\documentclass[sigconf,natbib=true]{acmart}
\usepackage{multirow}
\usepackage{arydshln}
\usepackage{graphicx}
\usepackage{subcaption} 
\usepackage{todonotes}
\usepackage{balance}
\usepackage{xcolor}

\AtBeginDocument{%
  }

\copyrightyear{2026}
\acmYear{2026}
\setcopyright{cc}
\setcctype{by}
\acmConference[SIGIR '26]{Proceedings of the 49th International ACM SIGIR Conference on Research and Development in Information Retrieval}{July 20--24, 2026}{Melbourne, VIC, Australia}
\acmBooktitle{Proceedings of the 49th International ACM SIGIR Conference on Research and Development in Information Retrieval (SIGIR '26), July 20--24, 2026, Melbourne, VIC, Australia}
\acmDOI{10.1145/3805712.3808552}
\acmISBN{979-8-4007-2599-9/2026/07}

\begin{document}

\title{RE-TRIANGLE: Does TRIANGLE Enable Multimodal Alignment Beyond Cosine Similarity in Retrieval?}


\author{Arijit Ghosh}
\authornote{All authors contributed equally to this research.}
\email{arijit.ghosh@student.uva.nl}
\orcid{0009-0003-5199-0809}
\affiliation{%
  \institution{University of Amsterdam}
  \city{Amsterdam}
  \country{The Netherlands}
}

\author{Aritra Bandyopadhyay}
\authornotemark[1]
\email{aritra.bandyopadhyay@student.uva.nl}
\orcid{0009-0009-6082-5593}
\affiliation{%
  \institution{University of Amsterdam}
  \city{Amsterdam}
  \country{The Netherlands}
}

\author{Chiranjeev Bindra}
\authornotemark[1]
\email{chiranjeev.bindra@student.uva.nl}
\orcid{0009-0007-5451-7135}
\affiliation{%
  \institution{University of Amsterdam}
  \city{Amsterdam}
  \country{The Netherlands}
}

\author{Jingfen Qiao}
\authornotemark[1]
\authornote{Corresponding author.} 
\email{j.qiao@uva.nl}
\orcid{0000-0002-4474-6213}
\affiliation{%
  \institution{University of Amsterdam}
  \city{Amsterdam}
  \country{The Netherlands}
}

\renewcommand{\shortauthors}{Arijit Ghosh, Aritra Bandyopadhyay, Chiranjeev Bindra, and Jingfen Qiao}
\begin{abstract}
Multimodal alignment is critical for bridging the semantic gap in information retrieval. However, traditional pairwise strategies introduce a geometric blind spot: while they align anchor modalities (e.g., text) with others, they lack constraints to enforce mutual consistency between peripheral modalities (e.g., video and audio). The TRIANGLE framework addresses this by minimizing the area of modality triplets on a hypersphere to enforce holistic alignment. In this reproducibility study, we verify the robustness of this geometric objective for retrieval tasks. We confirm that TRIANGLE outperforms pairwise baselines in zero-shot settings, achieving Recall@1 gains of up to +8.7 points, though benefits are domain-dependent. However, we fail to reproduce the reported learning-from-scratch results. Analysis using a synthetic toy dataset attributes this to instability when jointly optimizing geometric alignment with Data-Text Matching (DTM) loss. Furthermore, we find that cosine regularization primarily stabilizes text-to-video retrieval, and fine-tuning with domain supervision amplifies geometric benefits but reduces cross-dataset generalization. Our findings support the efficacy of geometric alignment while highlighting critical optimization sensitivities.\footnote{Code available at https://github.com/ARIJIT00171/RE-TRIANGLE}
\end{abstract}

\ccsdesc[500]{Information systems~Multimedia and multimodal retrieval}

\keywords{Multimodal retrieval, Contrastive Learning, Multimodal IR, Tri-modal alignment}


\maketitle

\section{Introduction}
\begin{figure}[ht!]
\includegraphics[width=\linewidth]{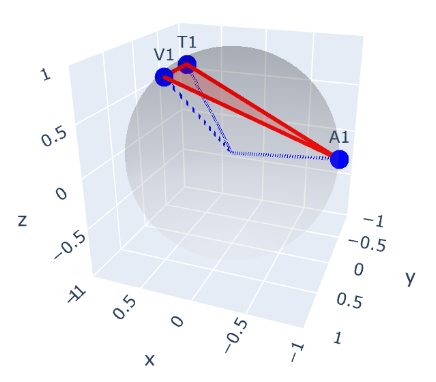}
\caption{TRIANGLE loss aims to minimize the Triangle area formed by the embedding vectors on a unit hypersphere between three modalities - Video (V1), Audio (A1) and Text(T1).}
\label{fig:overview}
\end{figure}
Multimodal alignment is a fundamental challenge in Multimodal Information Retrieval, where the goal is to bridge the semantic gap between multiple co-occurrent signals. This field was revolutionized by foundational dual-encoder models like CLIP~\cite{radford2021learning} and ALIGN~\cite{jia2021scaling}, which leverage pairwise contrastive learning to align images and text. Following this paradigm, numerous works have focused on refining alignment mechanics~\cite{uesaka2024, Zhai_2023_ICCV} or extending the architecture to specific modality pairs, such as audio-text in CLAP~\cite{elizalde2023clap}, video-text in CLIP4Clip~\cite{luo2022Neurocomput}, and point cloud-text in PointCLIP~\cite{zhang2022pointclip}. This approach has been scaled to more than two modalities by selecting a central anchor and aligning all other representations to it one by one, such as ImageBind~\cite{girdhar2023cvpr} and VAST~\cite{chen2023vast}. 

However, this pairwise alignment introduces a significant geometric blind spot~\cite{cicchetti2025triangle}. While the anchor modality aligns with others, there is no explicit constraint ensuring that peripheral modalities (e.g., Video and Audio) are effectively aligned with one another. Similarly, fusion-based methods~\cite{chen2023vast} reduce the tri-modal problem back to a pairwise objective (Fused Vector vs. Text), effectively masking the fine-grained geometric structure of the latent space. Recent efforts to model the joint distribution of $N$ modalities, such as Symile~\cite{saporta2024neurips} (optimizing total correlation) and GRAM~\cite{cicchetti2025icml} (minimizing Gramian volume), offer theoretical alternatives. Yet, these often suffer from computational instability or the "vanishing volume" problem in high dimensions, failing to outperform specialized baselines in tri-modal tasks.

To address this, \cite{cicchetti2025triangle} proposed TRIANGLE: TRI-modAl Neural Geometric LEarning. Rather than relying on independent 2D projections via cosine similarity, TRIANGLE computes a similarity measure directly in the higher-dimensional space spanned by three modality embeddings. This similarity is defined by the area of the triangle formed by the three modality embeddings lying on a unit hypersphere, replacing the standard cosine similarity metric. By minimizing this triangle area, the model encourages a holistic, joint alignment among all three modalities simultaneously. The authors claim this approach provides a reliable, interpretable metric for modality alignment and achieves state-of-the-art results, improving the performance of cosine-based methods by up to 9 points in Recall@1\\

Despite the theoretical elegance and reported performance gains of the TRIANGLE objective, its practical robustness and generalizability require independent verification. In this paper, we present a comprehensive reproducibility study motivated by three core research objectives:

\begin{itemize}
    \item \textbf{RQ1 (Alignment Effectiveness in Multimodal Retrieval):} Can a tri-modal geometric objective outperform independent pairwise alignment in multimodal retrieval tasks?

    \item \textbf{RQ2 (Out of domain performance):} Can the pre-trained tri-modal geometric objective effectively adapt to out-of-domain distributions through continuous fine-tuning?

    \item \textbf{RQ3 (Effectiveness of Learning from Scratch):}    Does tri-modal geometric objective also helps when training from scratch?

    \item \textbf{RQ4 (Interpretability of Triangle Similarity):} Does the triangle area provide a consistent and geometrically unambiguous indicator of modality alignment during training and inference?
\end{itemize}
In response to these questions, we present the following findings:

\begin{itemize}
    \item We corroborate the original finding that the tri-modal geometric objective outperforms independent pairwise alignment in zero-shot retrieval tasks across diverse datasets.

    \item We extend the evaluation to YouCook2~\cite{youcook2} to test if the tri-modal geometric objective holds up in out-of-domain video retrieval. We find that the objective is highly domain-sensitive, significantly underperforming in the unseen homogeneous YouCook2 domain compared to the baselines.  

    \item Investigating the reproducibility issues for the learning-the-space-from-scratch experiment by  curating a small toy dataset consisting of geometric shapes for a controlled study.

    \item Our qualitative analysis validates the Triangle area as an interpretable similarity metric. Our visualizations reveal that while text and video align closely, audio remains geometrically distant, explaining the model's uneven cross-modal performance .
    
\end{itemize}

\section{Overview of Triangle Alignment}
\subsection{Preliminaries: Pairwise Alignment}
Conventional multimodal retrieval relies on maximizing pairwise cosine similarity between an anchor (e.g., Text) and other modalities (e.g., Audio/Video). Formally, given a batch size \(B\) and a temperature parameter $\tau$, the symmetric pairwise contrastive loss between embeddings \(x\) and \(y\) is computed in Equation~\ref{eq:pairwise}.

\begin{equation} 
\label{eq:pairwise}
\mathcal{L}_{Pair} = -\frac{1}{2B} \sum_{(x,y)} \sum_{i=1}^{B} \log \frac{ \exp \left(\mathbf{x}_i^{\top}\mathbf{y_i}/\tau\right) }{ \sum_{j=1}^{B} \exp \left(\mathbf{x}_i^{\top}\mathbf{y}_j/\tau\right)}
\end{equation}

where the outer summation iterates over the directional pairs $(x,y) \in \{(\mathbf{a},\mathbf{t}), (\mathbf{t},\mathbf{a}), (\mathbf{v},\mathbf{t}), (\mathbf{t},\mathbf{v})\}$. The $\mathbf{t}$, $\mathbf{a}$, and $\mathbf{v}$ denote the normalized embeddings for text, audio, and video, respectively. In this setting, the peripheral relationships (e.g., Audio-Video) are not explicitly optimized. To address this, VAST aggregates video $\mathbf{v}$ and audio $\mathbf{a}$ representations into a unified vector via a MLP fusion layer, and the contrastive loss is applied between this single fused vector and the text. However, this approach collapses the distinct geometric structure of the individual modalities into a single point before alignment.

\subsection{Geometric Tri-modal Alignment}
To address the limitations of the independent pairwise alignment, TRIANGLE proposed to align three modalities concurrently by maintaining their distinct positions in the joint latent space, and posits that the area of this triangle serves as a superior similarity metric. Specifically, it utilizes three backbone encoders to map different modality inputs into $\mathbf{x}$, $\mathbf{y}$, and $\mathbf{z}$ embeddings of fixed dimensionality. These vectors are normalized to the unit norm, constraining them to lie on the surface of a unit hypersphere.

\textbf{Triangle Area Similarity}. As illustrated in Figure~\ref{fig:overview} (and further detailed in Figure 4), the extremities of the three embedding vectors define the vertices of a triangle. Let $\mathbf{u} = \mathbf{x} - \mathbf{y}$ and $\mathbf{w} = \mathbf{x} - \mathbf{z}$ are two triangle sides computed among
the three embeddings x, y, and z of the three modalities. The alignment similarity $A$ is derived from the triangle area, augmented by a pairwise regularization term, as shown in Equation~\ref{eq:area}. 

\begin{equation}
\label{eq:area}
A = \frac{1}{2}\sqrt{\langle \mathbf{u}, \mathbf{u} \rangle
\langle \mathbf{w}, \mathbf{w} \rangle
- \langle \mathbf{u}, \mathbf{w} \rangle^{2}}
- \alpha \cos \theta_{xy}
\end{equation}

where\(\langle \cdot , \cdot \rangle \) is the dot product. The first term represents the geometric area, which minimizes only when all three vectors are proximal. The second term is a regularization factor where $\cos \theta_{xy}$ is the cosine similarity between the specific modality pair $(x, y)$ in the retrieval task. This ensures that while the triplet geometry is optimized, the task-relevant pairwise alignment is explicitly encouraged preventing the optimization from drifting into geometric configurations that satisfy the area constraint but degrade pairwise retrieval.

\begin{table*}[h!]
\centering
\small
\caption{
\textbf{Zero-Shot Video Retrieval Results.} We report the original paper's Recall@1 (R@1) alongside our reproduced results for both Text-to-Video (T2V) and Video-to-Text (V2T) tasks. In addition to R@1, we report Recall@10, nDCG@10, and Reciprocal Rank@10 (RR) to assess ranking quality.
\textbf{$\Delta$} indicates the improvement of TRIANGLE (with $L_{DTM}$) over our reproduced VAST baseline.
$^{\ast}$ indicates statistically significant differences between TRIANGLE (with $L_{DTM}$) and VAST under a two-tailed paired permutation test ($p < 0.01$).
}
\setlength{\tabcolsep}{4pt}
\begin{tabular}{l  cc  cccc  cccc}
\toprule
\toprule
\multirow{3}{*}{\textbf{Dataset / Method}} & 
\multicolumn{2}{c}{\textbf{Original}} & 
\multicolumn{4}{c}{\textbf{Reproduced (Text $\rightarrow$ Video)}} & 
\multicolumn{4}{c}{\textbf{Reproduced (Video $\rightarrow$ Text)}} \\
\cmidrule(lr){2-3} \cmidrule(lr){4-7} \cmidrule(lr){8-11}
& R@1 (T2V)& R@1 (V2T)& R@1 & R@10 & nDCG@10& RR@10& R@1 & R@10 & nDCG@10& RR@10\\
\midrule

\multicolumn{11}{l}{\textit{\textbf{MSR-VTT}}} \\
COSINE & -- & -- & 44.8 & \textbf{82.1} & 62.7 & 56.5 & 38.1 & 77.7 & 56.8 & 50.3 \\
VAST  & 49.3 & 43.7 & 51.0 & 74.0 & 62.4 & 58.7 & 52.8 & 77.0 & 64.9 & 61.0 \\
TRIANGLE & \textbf{55.2} & \textbf{52.5} & \textbf{54.4}$^{\ast}$ & 81.3 & \textbf{67.5}$^{\ast}$ & \textbf{63.2}$^{\ast}$ & \textbf{54.6} & \textbf{82.7}$^{\ast}$ & \textbf{68.5}$^{\ast}$ & \textbf{64.0}$^{\ast}$ \\
\cdashline{1-11}
\textit{$\Delta$ w.r.t VAST} & +5.9 & +8.8 & +3.4 & +7.3 & +5.1 & +4.5 & +1.8 & +5.7 & +3.6 & +3.0 \\
\midrule

\multicolumn{11}{l}{\textit{\textbf{DiDeMo}}} \\
COSINE & -- & -- & 38.5 & 71.4 & 54.5 & 49.1 & 37.4 & 71.1 & 53.5 & 48.0 \\
VAST  & 49.5 & 48.2 & 50.6 & \textbf{78.2} & \textbf{64.5} & \textbf{60.0} & 51.4 & 78.5 & \textbf{64.9} & \textbf{60.6} \\
TRIANGLE & \textbf{54.9} & \textbf{53.1} & \textbf{51.7}$^{\ast}$ & 75.6 & 63.6 & 59.7 & \textbf{51.5}$^{\ast}$ & \textbf{78.9}$^{\ast}$ & 64.8 & 60.2 \\
\cdashline{1-11}
\textit{$\Delta$ w.r.t VAST} & +5.4 & +4.9 & +1.1 & -2.6 & -0.9 & -0.3 & +0.1 & +0.4 & -0.1 & -0.4 \\
\midrule

\multicolumn{11}{l}{\textit{\textbf{ActivityNet}}} \\
COSINE & -- & -- & 45.8 & 84.7 & 64.5 & 58.1 & 39.5 & 80.6 & 58.9 & 52.1 \\
VAST  & 51.4 & 46.8 & 49.8 & 81.0 & 65.1 & 60.0 & 46.7 & 75.0 & 60.6 & 56.0 \\
TRIANGLE & \textbf{59.7} & \textbf{54.1} & \textbf{58.5}$^{\ast}$ & \textbf{87.3}$^{\ast}$ & \textbf{72.8}$^{\ast}$ & \textbf{68.1}$^{\ast}$ & \textbf{58.8}$^{\ast}$ & \textbf{88.6}$^{\ast}$ & \textbf{73.5}$^{\ast}$ & \textbf{68.7}$^{\ast}$ \\
\cdashline{1-11}
\textit{$\Delta$ w.r.t VAST} & +8.3 & +7.3 & +8.7 & +6.3 & +7.7 & +8.1 & +12.1 & +13.6 & +12.9 & +12.7 \\
\midrule

\multicolumn{11}{l}{\textit{\textbf{VATEX}}} \\
COSINE & -- & -- & 75.2 & 98.1 & 87.9 & 84.5 & 69.9 & 97.4 & 84.5 & 80.3 \\
VAST  & 80.0 & 77.3 & 79.2 & 95.8 & 88.2 & 85.7 & 77.1 & 95.4 & 87.2 & 84.4 \\
TRIANGLE & \textbf{83.9} & \textbf{80.9} & \textbf{82.0}$^{\ast}$ & \textbf{98.1}$^{\ast}$ & \textbf{90.9}$^{\ast}$ & \textbf{88.5}$^{\ast}$ & \textbf{82.2}$^{\ast}$ & \textbf{98.7}$^{\ast}$ & \textbf{91.3}$^{\ast}$ & \textbf{88.9}$^{\ast}$ \\
\cdashline{1-11}
\textit{$\Delta$ w.r.t VAST} & +3.9 & +3.6 & +2.8 & +2.3 & +2.7 & +2.8 & +5.1 & +3.3 & +4.1 & +4.5 \\
\midrule

\multicolumn{11}{l}{\textit{\textbf{YouCook2}}} \\
COSINE & -- & -- & 11.2 & 39.5 & 23.6 & 18.8 & 8.7 & 32.3 & 19.0 & 14.9 \\
VAST  & -- & -- & \textbf{33.9} & \textbf{54.1} & \textbf{43.8} & \textbf{40.6} & \textbf{37.7} & \textbf{66.1} & \textbf{51.4} & \textbf{46.7} \\
TRIANGLE & -- & -- & 14.0 & 24.6 & 19.2 & 17.4 & 24.2 & 46.9 & 35.1 & 31.3 \\
\cdashline{1-11}
\textit{$\Delta$ w.r.t VAST} & -- & -- & -19.9 & -29.5 & -24.6 & -23.2 & -13.5 & -19.2 & -16.3 & -15.4 \\
\bottomrule
\bottomrule
\end{tabular}

\label{tab:text2video}
\end{table*}

\subsection{Tri-modal Optimization Objectives}
TRIANGLE adapts the standard contrastive objective by replacing the cosine similarity measure with the negative triangle area metric defined above. The loss function minimizes this area for positive triplets (pulling all three modalities together) while maximizing it for negative triplets. The optimization objective consists of two directional components: Data-to-Text ($\mathcal{L}_{D2T}$) and Text-to-Data ($\mathcal{L}_{T2D}$), where Data refers to the paired audio-video input.  


\begin{equation}
\label{eq:ld2t}
\mathcal{L}_{D2T}
= -\frac{1}{B}
\sum_{i=1}^{B}
\log
\frac{
\exp\!\left(-A(t_i, v_i, a_i)/\tau\right)
}{
\sum_{j=1}^{K}
\exp\!\left(-A(t_j, v_i, a_i)/\tau\right)
}
\end{equation}

\begin{equation}
\label{eq:lt2d}
\mathcal{L}_{T2D}
= -\frac{1}{B}
\sum_{i=1}^{B}
\log
\frac{
\exp\!\left(-A(t_i, v_i, a_i)/\tau\right)
}{
\sum_{j=1}^{K}
\exp\!\left(-A(t_i, v_j, a_j)/\tau\right)
}
\end{equation}

To further refine the alignment, a Data-Text Matching loss ($\mathcal{L}_{DTM}$) is employed. This binary classification objective predicts the probability $p_{dtm}$ that a text caption matches the corresponding audio-video features using a cross-attention mechanism:

\begin{equation}
\label{eq:ldtm}
\mathcal{L}_{DTM}
= \mathbb{E}_{(t,v,a)\sim (T,V,A)}
\big[
y \log p_{dtm}
+ (1 - y)\log(1 - p_{dtm})
\big]
\end{equation}

The final total loss $\mathcal{L}_{\mathrm{TOT}}$ is a weighted combination of the geometric contrastive losses and the matching loss:
 
\begin{equation}
\mathcal{L}_{\mathrm{TOT}}
= \frac{1}{2}
\left(
\mathcal{L}_{\mathrm{D2T}}
+ \mathcal{L}_{\mathrm{T2D}}
\right)
+ \lambda \mathcal{L}_{\mathrm{DTM}} .
\end{equation}

\section{Reproducibility and Replicability}
In this section, we address RQ1 and RQ2 by reproducing and replicating the experimental setup of the original study to verify its claims regarding the effectiveness of multimodal retrieval and learning efficiency. Firstly, we validate the model's multimodal retrieval effectiveness by evaluating it on the datasets mentioned in the original paper and extending the evaluation and training to a new dataset with different domain. Secondly, we assess its learning effectiveness and efficiency by training the model entirely from scratch. Additionally, while the original paper relied exclusively on Recall to measure performance, our study extensively evaluates ranking capability using broader metrics such as nDCG@10 and MRR@10. To ensure a strictly fair comparison, we also re-evaluate the VAST and a cosine similarity baseline on the exact same datasets used for the TRIANGLE experiments, which used backbones similar to TRIANGLE but used different alignment objectives and similarity measurement methods. The COSINE baseline in Tables~\ref{tab:text2video} and ~\ref{tab:audio2text} uses the same pretrained TRIANGLE encoders and computes retrieval scores via pairwise cosine similarity between text, audio, and video embeddings, without TRIANGLE area scoring. This isolates the effect of the similarity function, separating representation learning from geometric alignment.

\subsection{Replication of Multimodal Retrieval \label{sec:res_videoret}
Effectiveness}
\textbf{Experiment setup.} In this setting, we utilize the pre-trained TRIANGLE checkpoint released by the authors. This model uses BERT-Base~\cite{bertb} for text, BEATS~\cite{beats} for audio, and EVA-CLIP-ViT-G~\cite{eva} for video. These backbones were pre-trained on a subset of 150k video-audio-text triplets sampled from the VAST-27M dataset with TRIANGLE optimization objective. We evaluate this model across a diverse set of multimodal retrieval benchmarks, including video and audio retrieval. For video retrieval, we replicate the evaluation on MSR-VTT~\cite{msrvtt}, DiDeMo~\cite{didemo}, ActivityNet~\cite{activitynet}, and VATEX~\cite{vatex} datasets. To assess domain generalization, we extend the evaluation to YouCook2, a dataset of instructional cooking videos that serves as a stress test for the model. For audio retrieval, we utilize the AudioCaps~\cite{audiocaps} dataset and extend the original experimental scope to include the inverse audio-to-text task. The detailed statistics of all evaluation datasets are summarized in Table~\ref{tab:dataset_stats}.\\

\textbf{Claim 1: Effectiveness of Three-modal Alignment in Video Retrieval.} The video retrieval task comprises two subtasks: Text-to-Video (retrieving the most relevant video given a natural language query) and Video-to-Text (retrieving the most relevant caption given a video).

We successfully reproduced the results for both zero-shot video retrieval subtasks and extended the evaluation to a new dataset, YouCook2. Our results (Table \ref{tab:text2video}) largely confirm the main claims of the TRIANGLE paper. On MSR-VTT, ActivityNet, and VATEX, TRIANGLE consistently outperforms VAST across both retrieval directions and in all reported metrics. These results support the claim that jointly leveraging text, video, and audio through a tri-modal alignment objective can improve downstream retrieval performance compared to pairwise baselines. Notably, scores on ActivityNet are substantial, with improvements of up to +8.7\% Recall@1 for Text-to-Video retrieval and +12.1\% for Video-to-Text retrieval, indicating that the triangle area objective is particularly effective in aligning modality in diverse and long videos.

However, our extended analysis reveals that the effectiveness of geometric alignment is not universal. On DiDeMo, while TRIANGLE improves Recall@1, it shows negligible gains or slight drops in deeper ranking metrics (nDCG@10, RR@10). This indicates that while the objective effectively pushes the ground truth to the top, it does not uniformly improve the semantic ordering of the remaining candidates. 

Most critically, TRIANGLE significantly underperforms VAST on other domain of dataset YouCook2, suffering major drops across all metrics. In the zero-shot setting, TRIANGLE achieves only 14.0 R@1 (T2V) compared to VAST's 33.9. A similar degradation is observed for Video-to-Text retrieval (24.2 vs 37.7 R@1). Unlike MSR-VTT and ActivityNet, which contain diverse web-scale content, YouCook2 consists of short, instructional cooking videos with repetitive visual contexts (e.g., similar kitchen scenes) and verb-centric procedural language. This suggests that the learned tri-modal alignment geometry is sensitive to domain-specific semantics and may not preserve retrieval robustness under distribution shift.

To investigate whether this failure stems from domain mismatch rather than the objective design itself, we fine-tuned TRIANGLE on YouCook2 (Table \ref{tab:youcook2_finetuning_combined}). 
This domain-specific optimization yields substantial improvements over the zero-shot TRIANGLE model (pretrained on VAST subset), increasing Text-to-Video R@1 from 14.0 to 33.15 and Video-to-Text R@1 from 24.2 to 33.18.  This confirms that the tri-modal objective retains the capacity to adapt to instructional data when explicitly supervised.

However, this adaptation comes at a high cost in generalization. As shown in Table \ref{tab:youcook2_finetuning_combined}, fine-tuning on YouCook2 leads to severe degradation in zero-shot performance on MSR-VTT and DiDeMo. On MSR-VTT, Text-to-Video R@1 drops from 54.4 to 24.43 (29.97 points), while on DiDeMo it drops from 51.7 to 16.75 (34.95 points). The degradation is even more noticeable in deeper ranking metrics such as R@10 and nDCG@10. This suggests that domain-specific fine-tuning of the tri-modal geometric objective reorients the embedding space toward instructional semantics, disrupting previously learned cross-modal geometry.
 
These results reveal a trade-off between domain-specific adaptation and global alignment stability. While the tri-modal objective can specialize effectively to instructional data, it does not preserve previously learned cross-domain alignment structure.\\

\begin{table*}[h!]
\centering
\small
\caption{
\textbf{Effect of fine-tuning TRIANGLE on YouCook2.} The table evaluates both the in-domain and out-of-domain dataset. Absolute drops and improvements are computed with respect to the pretrained TRIANGLE model before fine-tuning.
}
\setlength{\tabcolsep}{5pt}
\begin{tabular}{l cccc cccc}
\toprule
\toprule
\multirow{2}{*}{\textbf{Method}} &
\multicolumn{4}{c}{\textbf{Text $\rightarrow$ Video (T2V)}} &
\multicolumn{4}{c}{\textbf{Video $\rightarrow$ Text (V2T)}} \\
\cmidrule(lr){2-5} \cmidrule(lr){6-9}
& R@1 & R@10 & nDCG@10 & RR@10 
& R@1 & R@10 & nDCG@10 & RR@10 \\
\midrule

\multicolumn{9}{c}{\textbf{In-Domain (YouCook2)}} \\\addlinespace[1ex]
TRIANGLE w/o $L_{DTM}$
& 9.17 & 39.53 & 22.32 & 17.07  & 10.10 & 43.34 & 24.80 & 19.10  \\

TRIANGLE with $L_{DTM}$
& \textbf{33.15} & \textbf{62.48} & \textbf{47.44} & \textbf{42.69}  & \textbf{33.18} & \textbf{63.79} & \textbf{47.96} & \textbf{42.96}  \\
\cdashline{1-9}
\textit{Impr. w.r.t pretrained}
& \textit{+19.15} & \textit{+37.88} & \textit{+28.24} & \textit{+25.29}  & \textit{+8.98} & \textit{+16.89} & \textit{+12.86} & \textit{+11.66}  \\

\midrule
\multicolumn{9}{c}{\textbf{Out-of-Domain (MSR-VTT)}} \\\addlinespace[1ex]
TRIANGLE with $L_{DTM}$ (After FT)
& 24.43 & 34.16 & 29.20 & 27.62
& 33.82 & 57.01 & 45.03 & 41.24 \\
\cdashline{1-9}
\textit{Abs. Drop (pts)}
& \textit{-29.97} & \textit{-47.14} & \textit{-38.30} & \textit{-35.58}
& \textit{-20.78} & \textit{-25.69} & \textit{-23.47} & \textit{-22.76} \\

\midrule
\multicolumn{9}{c}{\textbf{Out-of-Domain (DiDeMo)}} \\\addlinespace[1ex]
TRIANGLE with $L_{DTM}$ (After FT)
& 16.75 & 25.48 & 21.24 & 19.86
& 27.18 & 48.35 & 37.57 & 34.12 \\
\cdashline{1-9}
\textit{Abs. Drop (pts)}
& \textit{-34.95} & \textit{-50.12} & \textit{-42.36} & \textit{-39.84}
& \textit{-24.32} & \textit{-30.55} & \textit{-27.23} & \textit{-26.08} \\

\bottomrule
\bottomrule
\end{tabular}
\label{tab:youcook2_finetuning_combined}
\end{table*}

\textbf{Claim 2: Effectiveness of Three-modal Alignment in Audio Retrieval} To verify if the geometric benefits extend beyond video, we investigated audio-text and text-audio alignment using AudioCaps~\citep{audiocaps}.

We successfully reproduced the Text-to-Audio results (Table~\ref{tab:audio2text}). While the gap in Recall@10 is smaller than originally reported, TRIANGLE shows significant improvements in ranking quality (nDCG and RR) compared to VAST. Furthermore, our extension to the inverse task (Audio-to-Text) yields similar improvement margins (Table~\ref{tab:audio2text}). This symmetry is a crucial finding: it confirms that the TRIANGLE objective does not just over-fit to a specific direction (e.g., T$\rightarrow$A) but creates a genuinely holistic latent space where text, video, and audio are mutually accessible.

\textbf{Answers to RQ1 and RQ2}: Our findings confirm that a tri-modal geometric objective significantly outperforms independent pairwise alignment in zero-shot settings, provided the target domain exhibits sufficient multimodal diversity. However, this advantage is not universal; in specialized or highly homogeneous domains, geometric alignment may be superseded by fusion-based methods.

Furthermore, while the pretrained tri-modal objective can successfully adapt to out-of-domain distributions via explicit fine-tuning, this adaptation incurs a severe cost to generalization. The fine-tuned model suffers catastrophic forgetting, exhibiting a drastic collapse in zero-shot performance on previously strong and diverse benchmarks like MSR-VTT and DiDeMo. Consequently, although tri-modal geometric alignment demonstrates strong domain-specific adaptability, it fundamentally lacks the ability to preserve its global cross-modal geometry under continuous fine-tuning.

\begin{figure}[ht!]
\includegraphics[width=\linewidth]{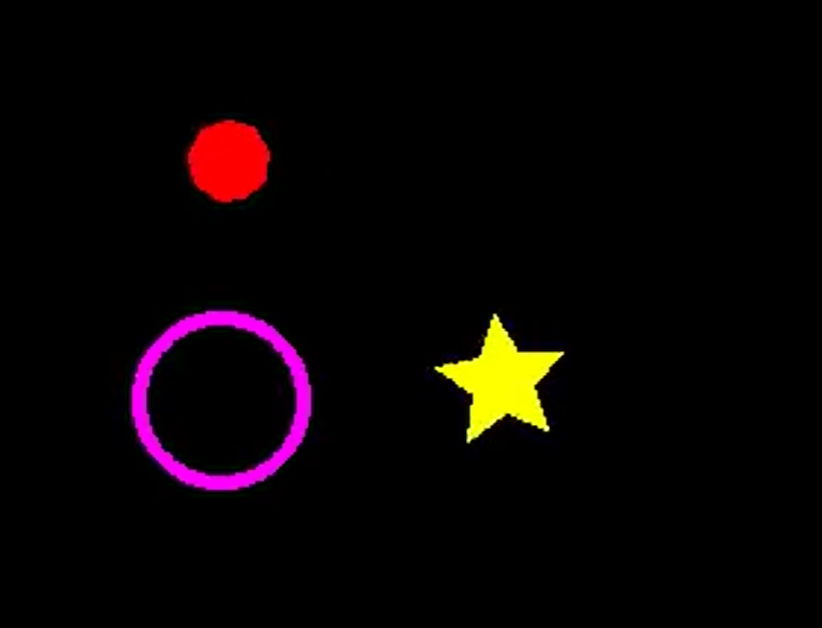}
\caption{Example of a video frame from the toy dataset consisting of video-audio-text triplets. Video consists of several colored shapes moving across a black background, and the text is a simple textual description ("This video contains 3 shapes: magenta ring, yellow star, red nonagon."). }
\label{fig:toydata}
\end{figure}

\subsection{Replication of Learning Effectiveness of Efficiency}
\textbf{Experimental Setup} We investigate the model's ability to learn the multimodal latent space without prior knowledge by initializing all encoders randomly. This is conducted in two controlled settings: We first train on the MSR-VTT dataset using the hyperparameter configuration (Table~\ref{tab:replication_hyperparameters}): batch size of 64, AdamW optimizer with a learning rate of $1e^{-4}$, and weight decay of $0.01$ for $79,249$ training steps. Inputs are sampled as 4 video frames and 1 audio frame per clip. To diagnose optimization stability, we also synthesize a controlled toy dataset of 9,000 aligned triplets as shown in Figure~\ref{fig:toydata}. This dataset comprises short video clips of moving geometric shapes, synthetic descriptive speech (without music or sound effects), and corresponding text subtitles. This simplified environment allows us to decouple the impact of the alignment loss from the complexity of real-world data.

\begin{table*}[h!]
\renewcommand{\arraystretch}{1.2}
\centering
\small
\caption{
\textbf{Zero-Shot Audio Retrieval Results on AudioCaps.} Comparison of Text-to-Audio (T2A) and Audio-to-Text (A2T) performance. We report the original paper's scores alongside our reproduced and extended results. TRIANGLE (with $L_{DTM}$) shows consistent improvements over VAST in both retrieval directions across most metrics. $\ast$ denotes statistically significant improvements of TRIANGLE over the \textbf{VAST} baseline
under a two-tailed paired permutation test ($p < 0.01$).
}
\setlength{\tabcolsep}{5pt}
\begin{tabular}{l  cc  cccc  cccc}
\toprule
\toprule
\multirow{3}{*}{\textbf{Method}} &
\multicolumn{2}{c}{\textbf{Original (Paper)}} &
\multicolumn{4}{c}{\textbf{Reproduced (Text $\rightarrow$ Audio)}} &
\multicolumn{4}{c}{\textbf{Extended (Audio $\rightarrow$ Text)}} \\
\cmidrule(lr){2-3} \cmidrule(lr){4-7} \cmidrule(lr){8-11}
& R@1 (T2A)& R@10 (T2A)& R@1 & R@10 & nDCG@10& RR@10& R@1 & R@10 & nDCG@10& RR@10\\
\midrule

COSINE & -- & -- & 26.75 & 66.94 & 45.29 & 38.54 & 25.65 & 66.26 & 44.78 & 38.07 \\
VAST & 32.1 & 65.4 & 25.65 & 71.47 & 47.08 & 39.48 & 29.22 & 73.25 & 49.95 & 42.62 \\

TRIANGLE
& \textbf{32.2} & \textbf{77.1}
& \textbf{32.92}$^{\ast}$ & 71.47 & \textbf{50.68} & \textbf{44.18}$^{\ast}$
& \textbf{32.37} & \textbf{74.90} & \textbf{51.88} & \textbf{44.72} \\
\cdashline{1-11}
\textit{Impr. vs VAST} & +0.1 & +11.7 & +7.27 & 0.00 & +3.60 & +4.70 & +3.15 & +1.65 & +1.93 & +2.10 \\

\bottomrule
\bottomrule
\end{tabular}

\label{tab:audio2text}
\end{table*}

\begin{table*}[h!]
\renewcommand{\arraystretch}{1.2}
\centering
\small
\caption{
\textbf{Learning from Scratch results on MSR-VTT.} Comparison of reported vs. reproduced results for retrieval in both directions (T2AV and AV2T). We experimented with two initializations of weights, one from the 150k VAST27M pre-trained checkpoint  (PT Weights) and the other without those weights. We observe that the former achieve scores consistent with those reported in the paper.
}
\setlength{\tabcolsep}{3.5pt}
\begin{tabular}{l  lcc  cccc  cc  cccc}
\toprule
\toprule
\multirow{3}{*}{\textbf{Method}}  &w/ PT Weights&
\multicolumn{6}{c}{\textbf{Text $\rightarrow$ Audio/Video (T2AV)}} &
\multicolumn{6}{c}{\textbf{Audio/Video $\rightarrow$ Text (AV2T)}} \\
\cmidrule(lr){2-8} \cmidrule(lr){9-14}

 && \multicolumn{2}{c}{Original} & \multicolumn{4}{c}{Reproduced} & \multicolumn{2}{c}{Original} & \multicolumn{4}{c}{Reproduced} \\
\cmidrule(lr){3-4} \cmidrule(lr){5-8} \cmidrule(lr){9-10} \cmidrule(lr){11-14}
 && R@1 & R@10 & R@1 & R@10 & nDCG & RR & R@1 & R@10 & R@1 & R@10 & nDCG & RR \\
\midrule

VAST
 &-& 36.5 & 79.3 & 1.36 & 17.65 & 7.58 & 4.63
& 35.5 & 77.3 & 0.90 & 15.05 & 6.36 & 4.55 \\

TRIANGLE w/o $L_{DTM}$
 &NO& - & - & 8.14 & 42.19 & 23.05 & 17.19
& - & - & 9.62 & 47.51 & 26.44 & 19.94 \\

TRIANGLE w/o $L_{DTM}$
 &YES& 33.3 & 74.4 & 29.86 & 74.2 & 50.21 & 42.74
& 40.4 & \textbf{81.7} & 36.65 & 80.09 & 57.21 & 50.03 \\
\cmidrule(lr){2-7} \cmidrule(lr){8-14}

TRIANGLE with $L_{DTM}$
 &NO& - & - & 1.81 & 18.44 & 8.07 & 5.03
& - & - & 1.58 & 19.68 & 8.54 & 5.26 \\

TRIANGLE with $L_{DTM}$
 &YES& \textbf{39.4} & \textbf{81.8} & \textbf{47.51} & \textbf{84.5} & \textbf{65.84} & \textbf{59.89}
& \textbf{41.9} & 80.0 & \textbf{47.51} & \textbf{84.16} & \textbf{65.69} & \textbf{59.81} \\

\bottomrule
\bottomrule
\end{tabular}

\label{tab:learn_from_scratch_table}
\end{table*}

\begin{table}[htbp]
\centering
\caption{Hyperparameter configuration for the replication of the learning from scratch experiments}
\label{tab:replication_hyperparameters}
\begin{tabular}{lll}
\hline
\textbf{Setting} & \textbf{MSR-VTT}  \\
\hline
Batch size & 64 \\
Optimizer & AdamW  \\
Learning rate & $1\times10^{-4}$ \\
Weight decay & 0.01  \\
Training steps & 79,249  \\
Video frames per clip & 4  \\
Audio frames per clip & 1  \\
Dataset size & 10000 \\
\hline
\end{tabular}
\end{table}

\textbf{Claim 3: Efficient Learning from Scratch}
\label{sec:res_leranfromscratch}
While the previous experiments focus on using pretrained models for evaluating retrieval performance, this experiment examines the ability of TRIANGLE to learn a multimodal embedding space entirely from scratch. All modality encoders are randomly initialized without any prior pretraining, allowing us to directly assess the model's capacity to learn cross-modal representations solely from the target dataset's distribution.

The original paper investigated this by training TRIANGLE from scratch on the MSR-VTT dataset for text-to-audio/video (T2AV) and audio/video-to-text (AV2T) retrieval tasks. The results of our reproduction attempts are shown in Table~\ref{tab:learn_from_scratch_table}. Our experiments found that the results align with the paper only if we start from the checkpoint (pretrained on a subset of 150k samples randomly selected from the VAST27M dataset). For standard pre-trained checkpoint initialization from HuggingFace, the scores are low, especially for the DTM module. This observation is consistent with the original paper, which mentions that the DTM is computed by passing visual and audio tokens through cross-attention layers in the text encoder (BERT-based model). Since BERT is an encoder-only model, these cross-attention layers are not present in the original HuggingFace checkpoint. They must be separately added, resulting in random initialization of the cross-attention module. This limits DTM performance when training from scratch and results in low scores.

To systematically investigate this reproduction gap and the negative impact of \(L_{DTM}\), we conducted a controlled experiment by training TRIANGLE and VAST from scratch on a synthetically generated toy dataset. The dataset was designed to be minimal yet multimodal, comprising video, audio, and text inputs. The video modality consisted of short clips depicting geometric shapes with varying colours and numbers of sides, moving and rotating across a plain background. The audio modality contained synthetic speech describing the visual content in a single sentence, while the text modality provided a corresponding textual description of the same scene. In total, we generated 9,000 aligned video–audio–text triplets to evaluate whether TRIANGLE and VAST could successfully learn joint embeddings from scratch in a simplified setting.

Retrieval results for this controlled setting are reported in Table~\ref{tab:toy_retrieval}. TRIANGLE trained without the \(L_{DTM}\) objective substantially outperforms both VAST and TRIANGLE variants that include \(L_{DTM}\). This suggests that the TRIANGLE loss alone is sufficient to learn a meaningful multimodal embedding space when training from scratch on simple datasets, whereas the inclusion of \(L_{DTM}\) may require further pretraining for the \(L_{DTM}\) loss to converge. Availability of the exact number of training steps used in the learning from scratch experiments from the original paper may help in reducing the \(L_{DTM}\) loss further.

\textbf{Answer to RQ3:} While the trimodal geometric objective does help in training the embedding space from scratch, the reproduced results fall significantly short of those reported in the original paper. Furthermore, the $L_{DTM}$ objective appears to hinder optimization in the absence of pretrained encoders. Therefore, while TRIANGLE demonstrates theoretical capacity for efficient learning from scratch, its practical effectiveness is strongly dependent on training stability and potentially undisclosed experimental details.

\begin{table*}[h!]
\centering
\small
\caption{
Toy dataset retrieval performance for learning from scratch using TRIANGLE and VAST. Notably, TRIANGLE without $L_{DTM}$ performs better than TRIANGLE with $L_{DTM}$.
}
\setlength{\tabcolsep}{12pt}
\begin{tabular}{lccccllll}
\toprule
\toprule
Method & \multicolumn{4}{c}{\textbf{Text $\rightarrow$ Video/Audio (T2AV)} }& \multicolumn{4}{c}{\textbf{Audio/Video $\rightarrow$ Text (AV2T)}}\\
\cmidrule(lr){2-5} \cmidrule(lr){6-9}
&R@1 &R@10 &nDCG@10 & RR@10  & R@1 & R@10 & nDCG@10 & RR@10  \\

VAST
& 3.4& 5.2& 4.4& 4.1& 3.0& 4.6& 3.91 & 3.67  \\

TRIANGLE w/o $L_{DTM}$
& \textbf{72.5}& \textbf{100.0}& \textbf{88.8}& \textbf{84.9}& \textbf{73.2}& \textbf{100.0}& \textbf{88.9}& \textbf{85.1}\\

TRIANGLE with $L_{DTM}$
& 0.1& 8.7& 3.2& 1.6& 0.1& 9.7& 3.6& 1.8\\
\bottomrule
\bottomrule
\end{tabular}
\label{tab:toy_retrieval}
\end{table*}

\begin{figure*}[t]
    \centering
    \begin{subfigure}[b]{0.48\linewidth}
        \centering
        \includegraphics[width=\linewidth]{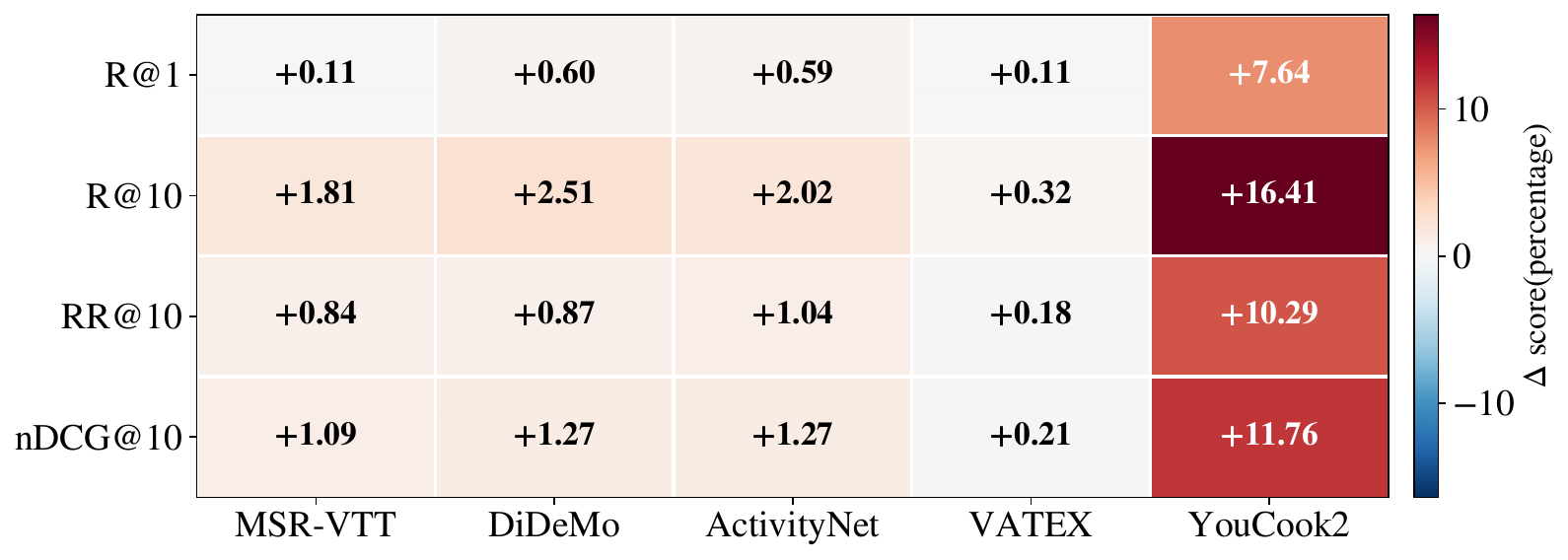}
        \caption{Text $\rightarrow$ Video retrieval}
        \label{fig:cosreg_t2v}
    \end{subfigure}
    \hfill 
    \begin{subfigure}[b]{0.48\linewidth}
        \centering
        \includegraphics[width=\linewidth]{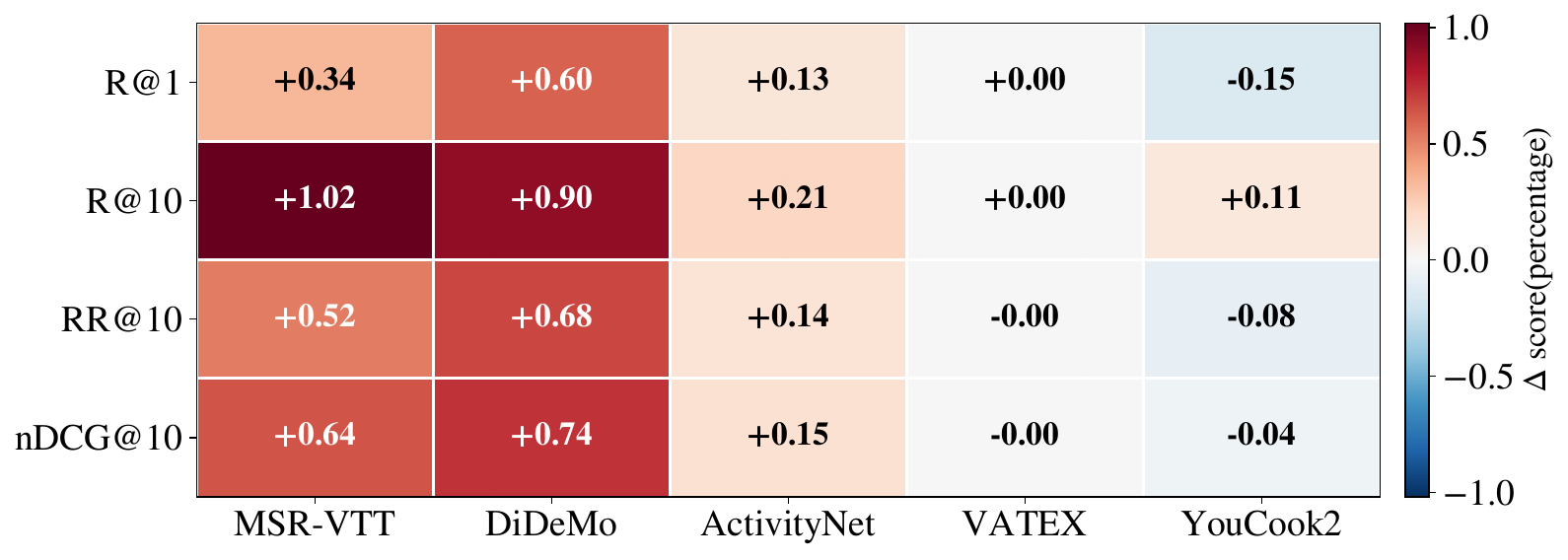}
        \caption{Video $\rightarrow$ Text retrieval}
        \label{fig:cosreg_v2t}
    \end{subfigure}
    
    \caption{Performance gain on adding cosine regularization ($\Delta =$ TRIANGLE with cosine regularization $-$ TRIANGLE without cosine regularization).}
    \label{fig:cosreg}
\end{figure*}

\section{Insights: Regularization and Interpretability}
Although our main results already include extensions such as additional ranking metrics and evaluation on YouCook2, this section focuses on two specific analyses that directly relate to claims made in the original TRIANGLE paper but are not fully supported by experimental evidence. First, we examine the effect of cosine regularization, which is described in the paper as an important component for stability in downstream retrieval tasks but is not explicitly implemented in the provided codebase. Second, we analyze the interpretability of the triangle-area metric, a property claimed by the authors but not supported visually through qualitative examples. These analyses are not intended to improve performance, but rather to examine how different components of the method influence retrieval outcomes.

\textbf{Effect of Cosine Regularization}
Figure~\ref{fig:cosreg} shows the effect of adding cosine regularization ($\alpha=1$) to the TRIANGLE objective for zero-shot video–text retrieval.

For Text-to-Video retrieval, cosine regularization consistently improves performance across most datasets, with particularly large gains on YouCook2 across all metrics. This indicates that constraining the angle between text and video embeddings is helpful when text is used to retrieve videos, as it reduces cases where the triangle area is small, even though the embedding vectors point in different directions. In contrast, for video-to-text retrieval, the effect of cosine regularization is much smaller. We observed only minor improvements on MSR-VTT, DiDeMo, and ActivityNet, and no noticeable changes on VATEX and YouCook2. This suggests that video embeddings already provide strong semantic information for retrieving text, with less angular inconsistencies. Overall, cosine regularization mainly improves text-to-video retrieval and acts as a stabilizing complement to the triangle-area objective.

\begin{figure*}[h!]
\centering
\begin{subfigure}{0.32\textwidth}
    \centering
    \includegraphics[width=\linewidth]{images/msrvtt_ex1.png}
    \caption{MSR-VTT positive example 1 ("spongebob is showing memories of him with mr") with area 1.02}
    \label{fig:msrvtt_ex1}
\end{subfigure}
\hfill
\begin{subfigure}{0.32\textwidth}
    \centering
    \includegraphics[width=\linewidth]{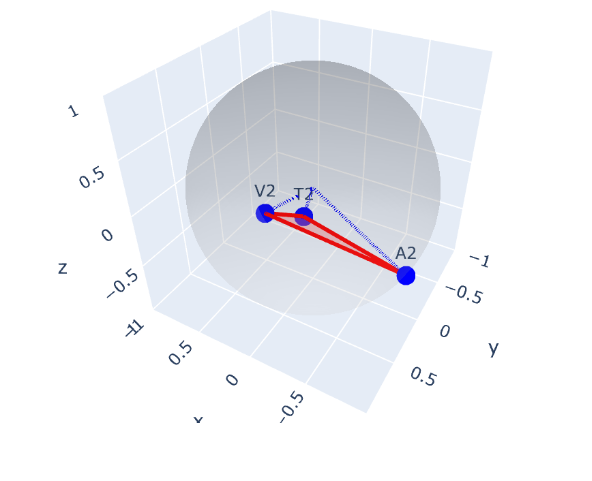}
    \caption{MSR-VTT positive example 2 ("a man in a flying contraption crashes in a field”) with area 0.91}
    \label{fig:msrvtt_ex2}
\end{subfigure}
\hfill
\begin{subfigure}{0.32\textwidth}
    \centering
    \includegraphics[width=\linewidth]{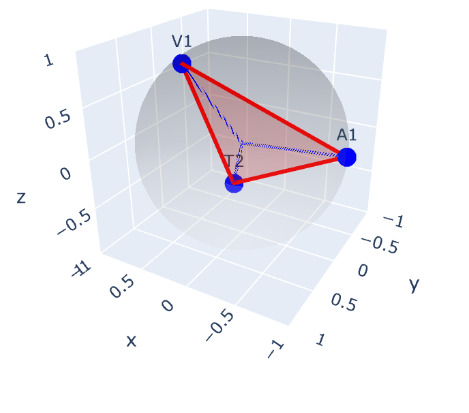}
    \caption{MSR-VTT negative example (text modality of positive example 1 swapped with that of positive example 2) with area 1.77}
    \label{fig:msrvtt_neg}
\end{subfigure}

\begin{subfigure}{0.32\textwidth}
    \centering
    \includegraphics[width=\linewidth]{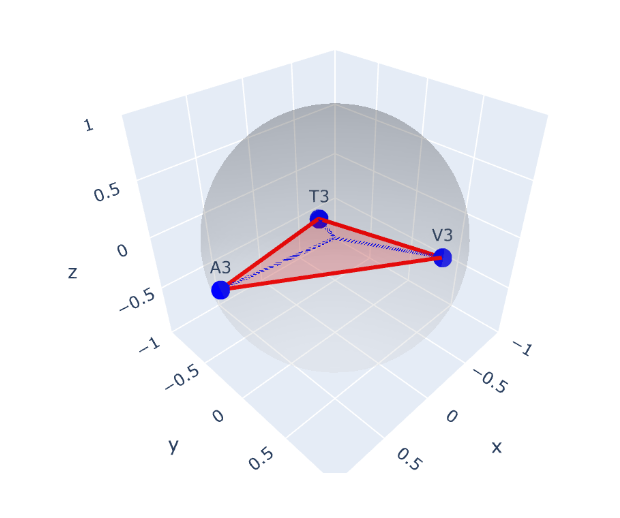}
    \caption{YouCook2 positive example 1 ("mash a bowl of chickpeas using a fork") with area 1.24}
    \label{fig:yc_ex1}
\end{subfigure}
\hspace{0.02\textwidth}
\begin{subfigure}{0.32\textwidth}
    \centering
    \includegraphics[width=\linewidth]{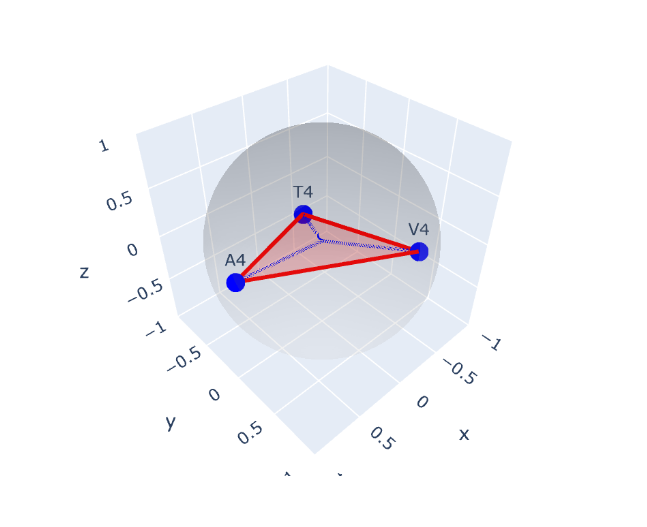}
    \caption{YouCook2 positive example 2 ("beat the dough on the table") with area 1.28}
    \label{fig:yc_ex2}
\end{subfigure}
\caption{Geometric interpretation of TRIANGLE-based multimodal alignment across datasets. Subcaptions state the video descriptions (text modality). Each plot visualizes the embeddings of Audio (A), Video (V), and Text (T) extracted from TRIANGLE’s backbone encoders, projected into 3D using PCA, and shown on a hypersphere. Small triangle areas indicate stronger multimodal alignment, while larger areas reflect weaker or misaligned modality representations. Negative examples have higher area values and are less aligned than positive examples. Positive examples from more complicated datasets like YouCook2 are less aligned than those from simpler datasets like MSR-VTT.}
\label{fig:explain}
\end{figure*}

\textbf{Geometric Interpretability of the Triangle Loss}
\label{sec:res_explain}
Claim 4 of the original paper states that TRIANGLE is an interpretable metric for modality alignment. However, the original paper does not include any qualitative examples or visualization diagrams to substantiate these claims. Our reproducibility study further investigates this claim by extracting the multimodal text, video, and audio vectors from the backbone encoders in TRIANGLE and visualizing them by reducing their dimensionality to 3 using PCA. The visualization results are shown in Figure~\ref{fig:explain}.

Figures~\ref{fig:msrvtt_ex1} and \ref{fig:msrvtt_ex2} show case multimodal vectors of two examples from the MSR-VTT dataset plotted on the hypersphere in 3D. Figure~\ref{fig:msrvtt_neg} showcases a negative example where the text modality of the first example is swapped with the text modality of the second example. As expected, the positive examples have a smaller area and are much more aligned than the negative example, which has a much larger area. Additionally, we observe that TRIANGLE effectively aligns the text and video modalities, as evidenced by their spatial proximity in the projected space. However, the audio embeddings remain comparatively distant from both text and video embeddings. This pattern suggests that, even on relatively structured datasets such as MSR-VTT, TRIANGLE encounters difficulty in jointly aligning the audio modality with the other modalities. The geometric visualization thus reveals modality-specific alignment disparities that are not apparent from retrieval metrics alone.

Figures~\ref{fig:yc_ex1} and \ref{fig:yc_ex2} show two positive examples from YouCook2 which is a much more complicated multimodal dataset. Interestingly, the positive examples of YouCook2 have a much larger triangle area than the positive examples of MSR-VTT. This mismatch across modalities suggests that TRIANGLE may face greater challenges in jointly aligning all modalities in more complex multimodal datasets.

\textbf{Answer to RQ4}: Yes, the triangle area provides a consistent and geometrically unambiguous indicator of modality alignment. Our interpretability experiments support RQ4 and allow us to delve deeper into the reasons for some of the shortcomings of TRIANGLE. But, it is important to note that the triangle area provides relative geometric interpretability per query rather than an absolute, universal threshold for relevance. The optimal area signifying alignment is inherently dataset-dependent, as evidenced by the larger baseline areas observed in complex instructional domains like YouCook2 compared to MSR-VTT.

\section{Discussion}
Our reproducibility results support the core claims made by the original authors. Claims 1 and 3 were strongly supported by the experiments presented in the original paper, and our study successfully reproduces these findings. In particular, we confirm that TRIANGLE establishes a new state-of-the-art for zero-shot video-text and audio-text tasks by ensuring holistic modality alignment. While Claim 2 was not properly substantiated with visual evidence in the original paper, our experiments in Section~\ref{sec:res_explain} validate that the triangle area provides an interpretable metric for modality alignment. Furthermore, we extend the original evaluation by reporting ranking-aware retrieval metrics such as nDCG and RR, which account for the ordering of retrieved items. Across these metrics, TRIANGLE consistently outperforms VAST, indicating that its multimodal alignment improves not only top-rank retrieval but also the overall relevance and quality of the ranked results.

\begin{table*}[t] 
\centering
\small
\caption{Statistical overview of the evaluation dataset.}
\setlength{\tabcolsep}{4pt} 
\begin{tabular}{@{} l r c c l l p{5.5cm} @{}} 
\toprule
\textbf{Dataset} & \textbf{Size} & \textbf{Avg. Length} & \textbf{Text Length} & \textbf{Source} & \textbf{Domain} & \textbf{Example Text Query} \\
\midrule
\multicolumn{7}{c}{\textbf{Video-Text Retrieval}} \\\addlinespace[1ex]
MSR-VTT      & 10,000 & ~14.5s & $\sim$9 words & Web videos & Open & ``A man extinguishes a fire outside.'' \\
ActivityNet  & 19,811 & $\sim$118s & $\sim$13 words & YouTube & Complex daily activities & ``They continue working out in the room.'' \\
DiDeMo       & 997 & ~27.5s & 131--169 chars & Flickr & Open (localized moments) & ``The glare of a bright light.'' \\
VATEX        & 944 & $\sim$15s & $\sim$15 words & Kinetics-600~\citep{kay2017kinetics} & Open & ``A young child receives a haircut from an adult.'' \\
YouCook2     & 2,736 & $\sim$5.3 min & $\sim$9 words & YouTube & Cooking (instructional) & ``Add oil and cook the egg.'' \\
\midrule
\multicolumn{7}{c}{\textbf{Audio-Text Retrieval}} \\\addlinespace[1ex]
AudioCaps    & 2,475 & $\sim$10s & $\sim$9 words & AudioSet~\citep{audioset} & Open & ``Someone has a hiccup while typing.'' \\
\midrule
\multicolumn{7}{c}{\textbf{Synthetic / Controlled Evaluation}} \\\addlinespace[1ex]
Toy Shapes   & 9,000 & 12s & $\sim$10 words & Synthetic & Moving colored shapes & ``This video contains 2 shapes: red star, blue circle.'' \\
\bottomrule
\end{tabular}
\label{tab:dataset_stats}
\end{table*}

However, our reproducibility work highlights critical limitations. We were unable to reproduce the results for the learning-from-scratch experiments in Section~\ref{sec:res_leranfromscratch} using the authors' original codebase. The inability to minimize the $\mathcal{L}_{DTM}$ during training suggests a potential optimization bottleneck or a high sensitivity to default hyperparameter configurations, which warrants deeper investigation. Moreover, VAST consistently outperformed TRIANGLE on the YouCook2 dataset for zero-shot video retrieval tasks (Section~\ref{sec:res_videoret}). This dataset-dependent behavior indicates that the effectiveness of triangle-based alignment is sensitive to domain characteristics not fully captured by Recall@1 alone. A plausible explanation is that YouCook2 consists exclusively of instructional cooking videos, resulting in a highly homogeneous domain where fine-grained visual cues are critical. In such settings, text descriptions and audio signals may only partially reflect the underlying visual content, making zero-shot geometric alignment particularly challenging.

In contrast, VAST employs fusion-based mechanisms that explicitly integrate complementary information across modalities. These learned interactions may help compensate for weak text and audio signals by providing rich semantics to the joint representation. This hypothesis aligns with our interpretability analysis (Section~\ref{sec:res_explain}), which demonstrates that TRIANGLE struggles to tightly align modalities on YouCook2. Together, these observations suggest that while geometric alignment is highly effective in diverse open-domain settings, fusion-based approaches may be more robust for specialized, domain-specific datasets in zero-shot scenarios.

Finally, our analysis of cosine regularization shows that it predominantly improves text-to-video retrieval in cases where the TRIANGLE area alone does not enforce complete alignment, while having minimal impact on video-to-text retrieval. This suggests that text queries, being inherently more abstract, benefit significantly more from explicit angular constraints than richer visual embeddings do.

\section{Related Works}
Multimodal retrieval has primarily been addressed using contrastive dual-encoder architectures that align pairs of modalities within a shared embedding space. CLIP~\citep{radford2021learning} and ALIGN~\citep{jia2021scaling} established this paradigm at scale by training image–text models with cosine-based InfoNCE objectives on large datasets. This formulation has since been extended to other modality pairs, including audio–text alignment in CLAP~\citep{elizalde2023clap}, video–text retrieval in CLIP4Clip~\citep{luo2022Neurocomput}, and point cloud–text alignment in PointCLIP~\citep{zhang2022pointclip}. Additional works have focused on refining alignment mechanisms and improving representation quality~\citep{uesaka2024, Zhai_2023_ICCV}. While highly effective, these methods fundamentally optimize pairwise similarity and do not explicitly enforce higher-order consistency across more than two modalities.

To scale contrastive learning to multimodal settings, anchor-based strategies have become common. Approaches such as ImageBind~\citep{girdhar2023cvpr} and VAST~\citep{chen2023vast} learn a unified embedding space by aligning multiple modalities to a shared anchor, typically text. This design enables flexible integration of heterogeneous inputs and has shown strong empirical performance. However, because each modality is aligned independently to the anchor, the relationship between non-anchor modalities (e.g., audio and vision) is only indirectly constrained.

Recent work has explored geometric formulations to explicitly model joint multimodal structure. Symile~\citep{saporta2024neurips} optimizes total correlation to capture higher-order dependencies across modalities, while GRAM~\citep{cicchetti2025icml} proposes minimizing the Gramian volume spanned by modality embeddings to encourage joint collapse in representation space. TRIANGLE~\citep{cicchetti2025triangle} adopts a related geometric perspective but specializes to tri-modal alignment, defining similarity as the area formed by three modality embeddings on a unit hypersphere. By directly minimizing this area, TRIANGLE enforces a single joint objective across all three modalities rather than decomposing the problem into independent pairwise constraints.

\section{Conclusion}
In this reproducibility study, we examined the TRIANGLE framework to verify its core claims. Our results confirm that triangle-based geometric alignment can improve retrieval performance over pairwise objectives in zero-shot retrieval settings. Importantly, our analysis also reveals crucial dataset- and direction-dependent effects, demonstrating that the benefits of tri-modal alignment are not uniform across all scenarios. Through comprehensive evaluations with ranking-aware metrics, cosine regularization analysis, and interpretability experiments, we provide a nuanced understanding of when and why TRIANGLE is effective. However, our study also highlights several limitations in the original codebase, most notably the inability to reliably reproduce the learning-from-scratch results reported in the original work. Overall, this work largely supports the efficacy of geometric alignment objectives for multimodal retrieval, while emphasizing the need for careful evaluation across diverse datasets and retrieval settings. Furthermore, evaluating geometric alignment frameworks on broader domains, such as visual document retrieval~\cite{qiao2025vdr, ma2024unifying, faysse2025colpali}, remains a promising direction.

\section*{Acknowledgments}
This work was initially developed in the context of the Information Retrieval 2 (IR2) course at the University of Amsterdam. We express our sincere gratitude to Dr. Mohammad Alian Nejadi and Dr. Panagiotis Efstratiadis for their guidance and support.

\bibliographystyle{ACM-Reference-Format}
\balance
\bibliography{reference}
\end{document}